\documentclass[aip, amsmath,amssymb, reprint,]{revtex4-2}

\usepackage{graphicx}
\usepackage{dcolumn}
\usepackage{bm}
\usepackage{amssymb} 
\usepackage{color,soul}
\usepackage{times,mathptmx}
\usepackage{amsmath}
\usepackage[normalem]{ulem}
\usepackage{soul}
\usepackage{etoolbox}
\usepackage[T1]{fontenc}
\usepackage{verbatim}

\newcommand{\edi}[1]{{\color{black}{#1}}}
\newcommand{\jb}[1]{{\color{black}{#1}}}

\makeatletter
\def\@email#1#2{%
 \endgroup
 \patchcmd{\titleblock@produce}
  {\frontmatter@RRAPformat}
  {\frontmatter@RRAPformat{\produce@RRAP{*#1\href{mailto:#2}{#2}}}\frontmatter@RRAPformat}
  {}{}
}%
\makeatother

\begin{document}
\title{Microfluidic free interface diffusion: measurement of  diffusion coefficients and evidence of interfacial-driven transport phenomena}
\author{Hoang-Thanh Nguyen}
\author{Anne Bouchaudy}
\author{Jean-Baptiste Salmon}
\email{jean-baptiste.salmon-exterieur@solvay.com}
\affiliation{CNRS, Solvay, LOF, UMR 5258, Univ. Bordeaux, F-33600 Pessac, France}

\date{\today}

\begin{abstract}
We have developed a microfluidic tool to measure the diffusion coefficient $D$ of solutes in an aqueous solution, by following  the temporal relaxation of an initially steep  concentration gradient in a  microchannel. 
Our chip exploits multilayer soft lithography and the opening of a pneumatic microvalve to trigger the interdiffusion of pure water and the solution initially separated in the channel by the valve, the so-called free interface diffusion technique.
Another microvalve at a distance from the diffusion zone closes the channel and thus suppresses  convection. Using this chip, we  have measured diffusion coefficients  of solutes in water with a broad size range, from small molecules to polymers and colloids, with values  in the range   
$D \in [10^{-13}- 10^{-9}]$~m$^2$/s.  The same experiments but with added colloidal tracers also revealed  diffusio-phoresis and diffusio-osmosis phenomena due to the presence of the solute concentration gradient. We nevertheless show that these interfacial-driven transport phenomena do not affect the measurements of the solute diffusion coefficients  in the explored concentration range.
  
\end{abstract}

\maketitle
\section{Introduction}

Mass transport by diffusion in a liquid mixture is often the rate-limiting step in many chemical processes and is therefore a key element  in their modeling and optimization~\cite{Cussler}. Classical techniques for measuring diffusion coefficients $D$ in liquid systems are based on the temporal monitoring of the transient relaxation of a concentration gradient in a diffusion cell using for instance interferometry~\cite{Creeth1955,Caldwell1957}  or Raman spectroscopy~\cite{Bardow2005}.  However, these measurements remain tedious and also difficult because  the slightest unwanted convection in the cell can dominate  mass transport by diffusion. 

It was  recognized early that microfluidic technologies could provide adequate tools to study  diffusion in liquids due to the small scales involved (10-100~$\mu$m)~\cite{Squires2005}. Many groups, for instance, performed accurate measurements of diffusion coefficients by measuring the mixing between two miscible coflowing  streams in a  channel, see, e.g., Refs.~\cite{Kamholz2001,Salmon2005} and  Ref.~\cite{Peters2017} for multicomponent  mixtures. 
These experiments are nevertheless  limited to dilute solutions (or small differences in concentration between the coflowing liquids) to avoid any coupling between flow and mass transport~\cite{Dambrine2009}.
These measurements are also often limited to  diffusion coefficients $D \geq 10^{-10}$~m$^2$/s, because  smaller $D$ would impose extremely small flow rates, and also small channel heights to avoid the  dispersion due to the Poiseuille flow~\cite{Ismagilov1999,Kamholz2000,Salmon2007}. Continuous on-chip electrophoresis has been developed to avoid the latter problem and thus access a wider range of $D$, but at the cost of being limited to electrolyte solutions~\cite{Estevez-Torres2007}.
Microfluidic pervaporation~\cite{Bouchaudy:18},  evaporation~\cite{Goehring2017,Roger2018} and ultra-filtration~\cite{Keita2021} have been shown to  overcome these limitations and provide diffusion coefficients  even in concentrated  solutions,  down to $D \simeq 10^{-12}$~m$^2$/s. To estimate~$D$, these experiments exploit measurements of solute concentration profiles resulting from the balance between   molecular diffusion and  advection by the flow induced by pervaporation, evaporation or ultra-filtration. However, these experiments require precise flow measurements   and specific experimental configurations.

Few microfluidic methodologies have been developed to measure $D$ from the temporal relaxation of a solute concentration gradient in a microchannel in the absence of convection, as in a classical diffusion cell.
In particular, 
Culbertson {\it et al.}  measured the diffusion coefficients of different solutes by following the time evolution of a  solute plug generated 
by  electrophoresis in a microfluidic channel~\cite{Culbertson2002}. Vogus~{\it et al.} also measured solute diffusion coefficients by following the propagation of a diffusion front in a  channel closed at one end by a  hydrogel membrane to suppress convection~\cite{Vogus2015}.
Microfluidic experiments that generate a steep concentration gradient in a microchannel in the absence of flows, as in conventional diffusion cell measurements, are even more rare.
Huebner {\it et al.}, for instance, used electrocoalence of a pair of microdroplets immobilized in a microchannel to trigger and follow the reaction-diffusion process between these two micro-reservoirs of solutes~\cite{Huebner2011}.  Yamada {\it et al.} developed an ingenious technique of removable walls in a microfluidic chamber to generate a  steep  concentration gradient between a solution and its solvent and follow its evolution in a static configuration\cite{Yamada2016}. More recently, Hamada {\it et al.} used the technique of coflowing  liquids in a  channel to  generate a concentration gradient by stopping the flow using external valves connected to the chip~\cite{Hamada2021}. This technique makes it possible to measure  diffusion coefficients down to $D \simeq 10^{-12}$~m$^2$/s, mostly estimated from the long-time relaxation of the concentration gradient, when the finite size of the microfluidic channel comes into play.

In a different context, Hansen {\it et al}. developed microfluidic chips integrating pneumatic microvalves, commonly called Quake valves~\cite{Unger2000}, to trigger the interdiffusion between a protein solution and  precipitants initially separated by the valve in a channel~\cite{Hansen2002}. This technique, referred to as  {\it free interface diffusion}~\cite{Salemme1972}, has been used to screen protein crystallization conditions at a high throughput and grow protein crystals in a convection-free environment~\cite{Hansen2003}. 
In the present work, we design a  chip following the concept of  microfluidic  free interface diffusion proposed by Hansen {\it et al.}~\cite{Hansen2002} that allows the generation of a steep solute concentration gradient in a closed microchannel at a specific time. The monitoring of the transient relaxation of this gradient allows us to measure diffusion coefficients $D$ of different solutes, over a wide range, 
$D \in [10^{-13}- 10^{-9}]$~m$^2$/s, due to the absence of convection.  
Beyond these measurements, we also evidence diffusio-phoresis (DP) and diffusio-osmosis (DO) phenomena~\cite{Derjaguin1972,Anderson1989,Marbach2019} by adding colloidal tracers in the microchannel. However, we show that these interfacial transport phenomena due to the presence of the solute concentration have no impact on the measurements presented.

\section{Experimental section}

\subsection{Chemicals and fluorescence measurements}
\begin{figure}[htbp]
    \centering
    \includegraphics{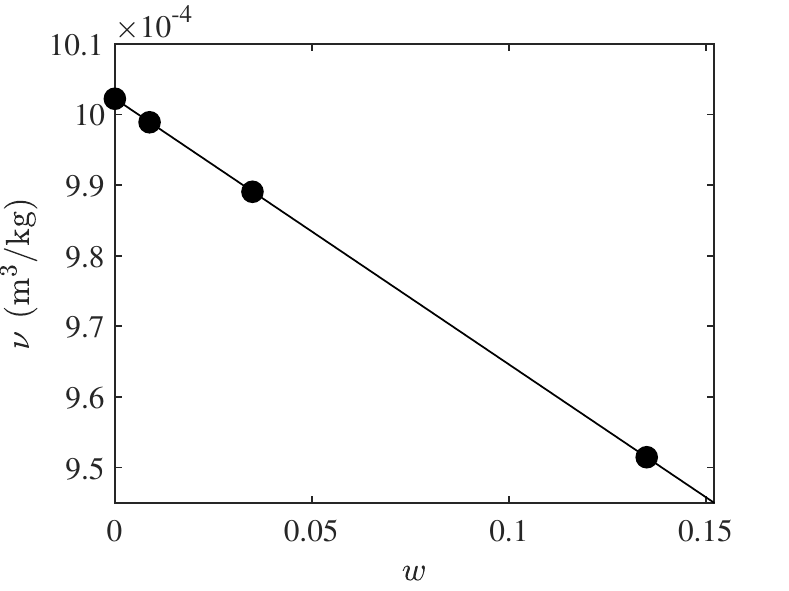}
    \caption{Specific volume $\nu$ of dextran 70~kDa solution against its mass fraction $w$. The continuous line is a fit by Eq.~(\ref{eq:nu}).
    \label{fig:SpecifVolume}}
\end{figure}

We studied  aqueous solutions of fluorescein  (Sigma Aldrich) at typical concentrations of 0.03~mM and fluorescein isothiocyanate dextrans (Sigma Aldrich) with molecular weights $M_w = 4$, 10, 20, and 70~kDa designated below as FD4, FD10, FD20, and FD70 respectively. 
In the latter case, the solutions were formulated by weighing with de-ionized water to achieve mass fractions  $w < 0.01$. 
We also measured the specific volume $\nu$ (m$^3$/kg)   of aqueous solutions of dextran 70~kDa  at various mass fractions and $T = 22^\circ$C, see Fig.~\ref{fig:SpecifVolume} (Anton-Paar, DMA 4500). 
These data are well-fitted in the range   $w=0$--$0.135$ by: 
\begin{eqnarray}
\nu =  \nu_w (1-w) +  \nu_d w\,, \label{eq:nu}
\end{eqnarray}
with $\nu_w =   1.002 \times 10^{-3}$~m$^3$/kg the specific volume of water at $T= 22^\circ$C and $\nu_d = 6.26 \times 10^{-4}$~m$^3$/kg. This linear behavior shows that the volume of the water/dextran  mixture does not change
significantly during mixing. The fitted value $\nu_d$ is in agreement with other values previously reported~\cite{Gaube1993} and was used to convert mass fractions $w$ of the FD solutions to  volume fractions $\varphi$ using  $\nu \varphi = \nu_d w$.

We also studied  aqueous dispersions of fluorescent monodisperse polystyrene  particles stabilized by sulfate groups (FluoSpheres, ThermoFisher Scientific)  either to track their trajectories in  Sec.~\ref{sec:evidenceof} (colloid diameter $2a = 2~\mu$m) or to measure their diffusion coefficient in Sec.~\ref{sec:colloiddisp} ($2a = 500$~nm). In both cases, the dispersions were diluted sufficiently to follow the trajectories of individual particles using particle tracking algorithms~\cite{PTcode}.

Concentration profile  measurements and tracking of colloidal particles  were performed using an inverted microscope (Olympus IX83) coupled to a sCMOS camera (OrcaFlash, Hamamatsu). 
To avoid significant photobleaching in the case of the fluorescein and FD solutions, we used a fluorescence illuminator with a shutter (X-Cite, Xylis) to limit the overall exposure during a complete experiment to a few seconds.

\subsection{Microfluidic chip}

\begin{figure}[htbp]
    \centering
    \includegraphics{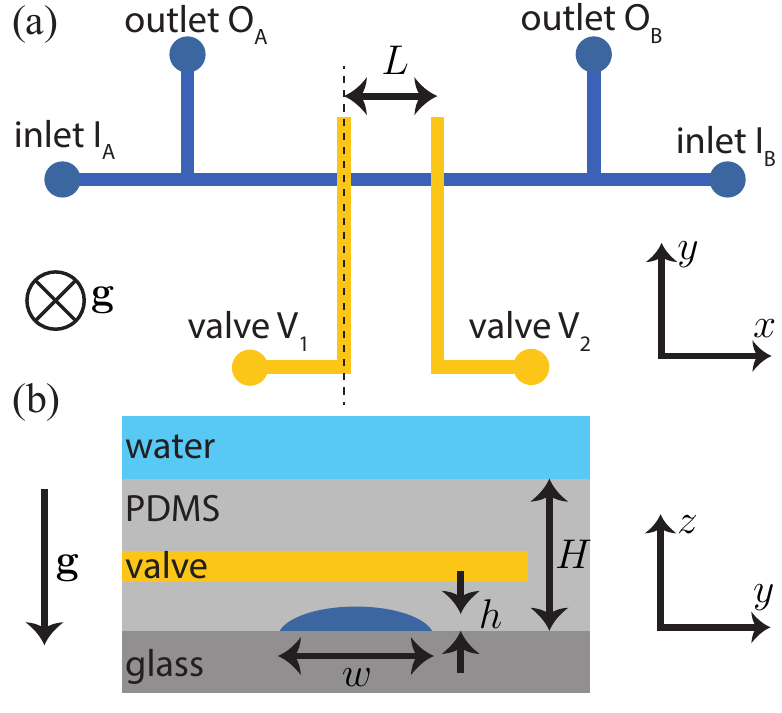}
    \caption{(a) Schematic view of the fluidic channel network (blue) 
		and the valves \jb{V$_1$ and V$_2$} (orange). \jb{The fluidic channel is composed of two inlets, I$_\text{A}$ and I$_\text{B}$, and two outlets, O$_\text{A}$ and O$_\text{B}$.} In the experiments reported, the distance between the two valves is either 
    $L=1$ or $L=4$~mm.
    (b) Schematic cross-section of the multilevel   chip \jb{along the dotted line shown in (a). To prevent pervaporation through the PMDS matrix, the chip is} immersed in a water bath.
    The dimensions are
    $h=23~\mu$m, $w=200~\mu$m, and $H \simeq 5$~mm. $\mathbf{g}$ is the gravity field.
    \label{fig:Setup}}
\end{figure}

Figure~\ref{fig:Setup} shows a schematic view of the chip we developed using  multilayer soft lithography to integrate pneumatically actuated valves~\cite{Unger2000}. 
In brief, the chip is composed of two layers of poly(dimethylsiloxane) (PDMS)  bonded to a glass slide using a plasma treatment.  The first layer contains rounded channels
(height $h \simeq 23~\mu$m, width $w \simeq 200~\mu$m) obtained by post heating a mold made using a positive photoresist (AZ-40XT, AZ Electronic Materials). The top PDMS layer contains two actuation rectangular channels (height 85~$\mu$m, width $200~\mu$m) and is obtained from a negative photoresist mold (SU8, MicroChem). In the following experiments, the center to center distance between the two valves is either $L = 1$~mm or $L=4~$mm. The dead-end actuation channels are filled by fluorinated oil (Fluorinert FC-40)  and   connected to a pressure controller (Fluigent, MFCS-EZ). 
The channels containing the fluids are separated from the actuation channels by a thin layer of PDMS ($\simeq 60~\mu$m) and the pressure required to  completely close the valves is $\simeq 1.6$~bar.

\subsection{Microfluidic protocol}

\begin{figure}[htbp]
    \centering
    \includegraphics{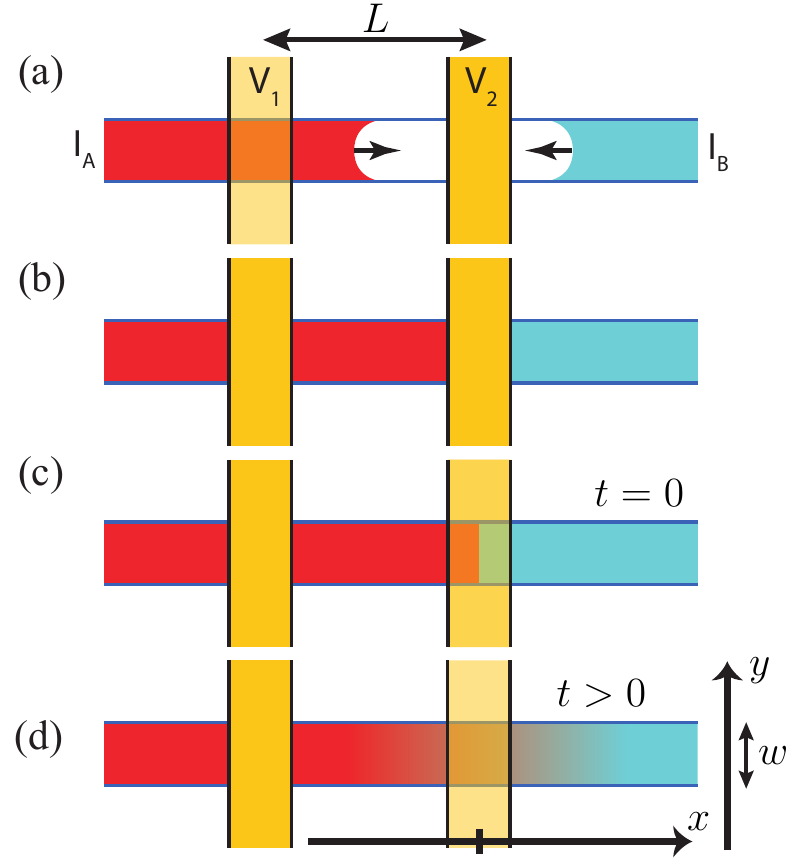}
    \caption{Protocol for the diffusion experiments.
    Open valves are indicated by the transparency of the orange color indicating them.
    (a) The solution is  injected from inlet I$_\text{A}$ and pure water from I$_\text{B}$, while valve V$_2$ is closed.
    (b) Valve V$_1$ is then closed and valve V$_2$ is opened at $t=0$ (c).  
    (d) Diffusion in the channel closed by valve V$_1$ ultimately smoothes the concentration gradient.   \label{fig:Protocol}}
\end{figure}

Figure~\ref{fig:Protocol} schematically shows  the protocol used to measure the  diffusion coefficient of solutes in an aqueous solution.
In a first step, valve V$_2$ is closed and the aqueous solution under study at concentration $\varphi_0$ and pure water are injected in the chip through inlets~I$_\text{A}$ and I$_\text{B}$ respectively, 
at an imposed pressure of $\simeq 300~$mbar (Fluigent, MFCS-EZ).  As soon as liquids flow out of outlets O$_\text{A}$ and O$_\text{B}$, these are manually closed by small plugs and   the gas permeability of PDMS  allows the two compartments of the main channel separated by valve V$_2$  to be filled by both the solution  and water, Fig.~\ref{fig:Protocol}(a). 
Once the channel is filled, the pressure at the inlets is released and valve V$_1$ is closed, Fig.~\ref{fig:Protocol}(b).
Valve~V$_2$ is then opened, which leads to a steep concentration gradient between the solution and water at a well-defined time $t=0$, Fig.~\ref{fig:Protocol}(c).
Diffusion then eventually smoothes this concentration gradient at later times in the channel closed by valve~V$_1$, Fig.~\ref{fig:Protocol}(d).

\begin{figure}[htbp]
    \centering
    \includegraphics{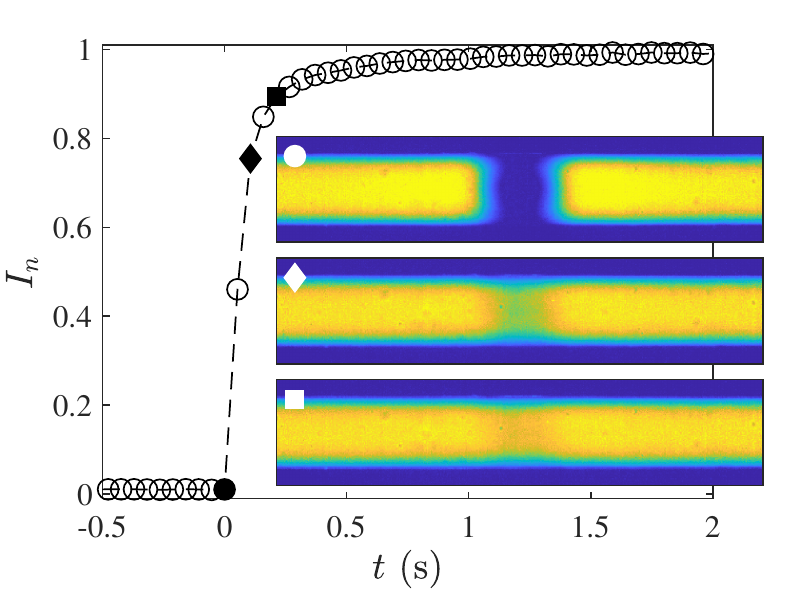}
    \caption{Opening dynamics of valve V$_2$. Both compartments are  filled with the same dilute fluorescein solution. $I_n$ is the normalized fluorescence intensity under the valve. The snapshots at $t=0$, $106$ and $212$~ms show the rapid opening of the valve. \label{fig:Valve}}
\end{figure}

To evaluate the characteristic  opening time  of valve~V$_2$, we used the protocol described in Fig.~\ref{fig:Protocol} but with the same dilute fluorescein solution to fill the channel. The opening of valve~V$_2$  is then  monitored using fluorescence microscopy at  high frame rate, Fig.~\ref{fig:Valve}. The analysis of the fluorescence measured under the valve allows to estimate  the opening time of this valve, $\tau \simeq 200~$ms for an opening at $\simeq 80\%$.
This characteristic time  is both related to the elasticity of the PDMS matrix and to the hydrodynamic resistance of the channel into which the volume of liquid left by the opening of valve~V$_2$ transiently flows~\cite{Goulpeau2005}.  
The rapid opening of the valve thus makes it possible to obtain a very steep initial solute concentration gradient using the protocol shown in Fig.~\ref{fig:Protocol}. Indeed,  molecular diffusion smoothes the concentration gradient during the opening of the valve at best on a length scale $\simeq \sqrt{D \tau} \simeq 15~\mu$m for a molecular solute with a diffusion coefficient $D =  10^{-9}$~m$^2$/s in water.

\subsection{Removal of pervaporation and role of free convection \label{sec:pervapobuo}}
In the protocol discussed above,  the closed valve V$_1$ prevents a priori any  convection in the channel and the mixing between the solution and water is thus purely diffusive. In fact, it is difficult to completely suppress all convection even in a dead-end channel because water pervaporation through the PDMS matrix inevitably induces flows in the channel~\cite{Verneuil:04,Randall:05,Bacchin2022}.  
To illustrate this point, the water pervaporation rate $q$ (m$^2$/s) per unit length 
can be estimated using an approximate relation derived in the case of a rectangular channel~\cite{Dollet2019}.
This calculation leads to $q \simeq 3\times 10^{-13}$~m$^2$/s for the geometric dimensions given in Fig.~\ref{fig:Setup} and an ambient  relative humidity RH$=0.45$. 
The characteristic time for emptying the channel by pervaporation $\tau_p = (hw)/q$ is a few hours and  not negligible compared to the relaxation time of the concentration gradient in some of the experiments discussed in this work.
In order to completely eliminate pervaporation leading not only to  undesirable flows, but also to the continuous accumulation by this flow of any traces of chemical  species contained in the  channel~\cite{Bacchin2022}, the  chip is immersed in a water bath during the experiments and also a few days before to saturate the PDMS matrix with water, see Fig.~\ref{fig:Setup}(b)~\cite{Verneuil:04,Randall:05}.

Furthermore, the difference in density between the solution and the water in
the experiment shown in Fig.~\ref{fig:Protocol}(d) also inevitably generates a flow because the density gradient is orthogonal to the gravity field $\mathbf{g}$. This solutal free convection flow then advects the solute and can influence the diffusive mixing between the solution and its solvent~\cite{Squires2005}.  This problem has been studied recently for both the case of a microfluidic slit and a channel of  rectangular cross-section~\cite{Salmon2021}. This theoretical work shows that the influence of free convection remains totally negligible for  Rayleigh numbers $\text{Ra} < 10^3$, with $\text{Ra}$ defined by: 
\begin{eqnarray}
\text{Ra} = \frac{\delta \rho  g  h^3}{\eta_w D}\,,
\end{eqnarray}
 $\delta \rho$ (kg/m$^3$) being the density difference between water and the studied solution and $\eta_w$  the water viscosity.
In the experimental conditions explored in this work, $\text{Ra} \ll 10^3$, and free convection plays no role on the interdiffusion between the solution and its solvent, although the microfluidic channel is horizontal and thus orthogonal to the gravity field~$\mathbf{g}$.

\section{Results}

\subsection{Molecular solutions: fluorescein and dextrans \label{sec:molsol}}

\begin{figure}[htbp]
    \centering
    \includegraphics{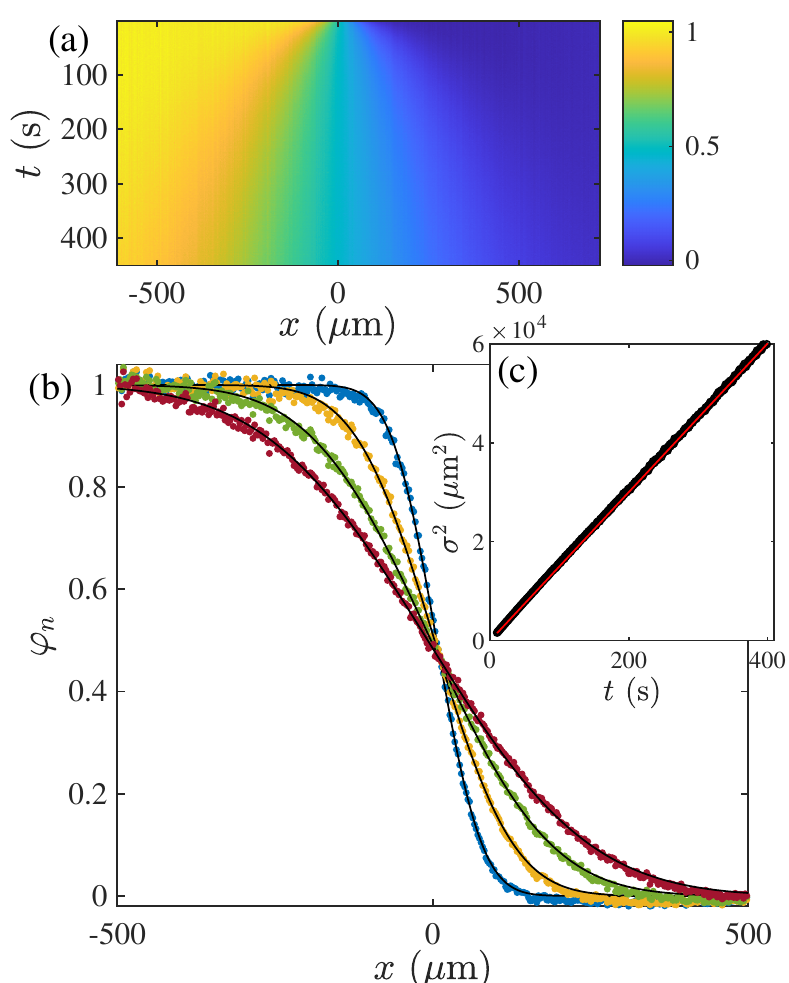}
    \caption{Dextran diffusion. (a) Space-time plot of the  normalized fluorescence profiles $\varphi_n(x,t)$ during the mutual diffusion of water and a FD4 solution \jb{(temporal resolution: one profile per second)}. (b) $\varphi_n(x,t)$ at $t=10$, 30, 70, and 130~s. The thin dark lines are fits by Eq.~(\ref{eq:eqDiff}).
    (c) $\sigma^2$ vs $t$ obtained from the fits of the data. The linear fit (red line) gives an estimate of the diffusion coefficient~$D$.
    \label{fig:FD4}}
\end{figure}

We first applied the methodology described above in the case of dilute aqueous solutions of molecules: fluorescein and fluorescent dextrans of different molecular weights.
All the measurements were performed at room temperature $T = 22^{\circ}$C. 
Figure~\ref{fig:FD4} shows the results for the dextran solution of molecular weight $M_w = 4~$kDa, referred below to as FD4, the polymer volume fraction of the injected solution being $\varphi_0 \simeq 0.005$ (estimated using the measured specific volumes shown in Fig.~\ref{fig:SpecifVolume}). 

Figure~\ref{fig:FD4}(a) shows the space-time plot of the fluorescence intensity profiles measured along the channel as a function of time from the opening of valve~V$_2$, see Fig.~\ref{fig:Protocol}(d) for the definition of the $x$ axis.
To obtain these profiles, the fluorescence intensity was corrected by the dark value obtained when the channel is filled with water, corrected by the inhomogeneity of the UV lamp (measured when the entire channel is filled by the FD4  solution), and normalized to 1 corresponding to the intensity of the initial FD4 solution at concentration $\varphi_0$. 
This space-time plot clearly shows the mutual diffusion  of the polymer solution and its solvent initially  separated by valve~V$_2$ in the channel.

In such experiments,   mass transport is a priori only governed by molecular  \jb{Fickian} diffusion because the microfluidic channel is closed by valve~V$_1$ and pervaporation as well as solutal free convection are negligible, see Sec.~\ref{sec:pervapobuo}. \jb{The volume average velocity 
of the binary mixture FD4 $+$ water is therefore zero, and} 
the polymer volume fraction  \jb{follows  the diffusion equation which is in this 1D geometry~\cite{Cussler}:
\begin{equation}
\frac{\partial \varphi}{\partial t}  = \frac{\partial }{\partial x}\left( D \frac{\partial \varphi}{\partial x}\right),
 \label{eq:eqDiffOrig}
 \end{equation}
where $D$ is the mutual diffusion coefficient of the mixture}, constant for a dilute solution. 
\jb{This equation implies that the polymer mass transport is solely due to molecular diffusion, and thus rules out any other mechanism, such as stress gradient-induced polymer migration, see for instance~\cite{Hajizadeh2017,Xiang2021}.
This hypothesis is very likely because there is no flow in the microfluidic channel closed by  valve V$_1$, and because it is unlikely that significant mechanical stresses were stored during the  filling step described in Fig.~\ref{fig:Protocol} and were not  released before the opening of valve V$_2$.}

\jb{For the initial condition, $\varphi(x<0,t=0) =\varphi_0$ and $\varphi(x>0,t=0) = 0$, the solution of Eq.~(\ref{eq:eqDiffOrig}) is~\cite{Crank}:}
\begin{equation}
 \varphi(x,t) = \frac{\varphi_0}{2} \left[1-\text{erf}\left(\frac{x}{2\sigma}\right)\right],
 \label{eq:eqDiff}
 \end{equation}
  where  $\sigma = \sqrt{D t}$ is the width of the diffusion zone.  
For this result, we implicitly assume an infinite geometry along $x$, and   Eq.~(\ref{eq:eqDiff}) is in fact only valid for $t\ll L^2/D$, where $L$ the is distance between the two valves ($L^2/D \simeq 4$~h for $L=4~$mm and $D=10^{-9}$~m$^2$/s). 
Figure~\ref{fig:FD4}(b) shows that $\varphi_n=\varphi(x,t)/\varphi_0$ given by Eq.~(\ref{eq:eqDiff}) perfectly fits the measured normalized intensity profiles validating the above assumptions and further showing that the fluorescence intensity varies linearly with $\varphi$ under these dilute conditions.
The linear behavior $\sigma^2$ vs\ $t$ shown in Fig.~\ref{fig:FD4}(c) leads to  $D \simeq 1.6 \times 10^{-10}~$m$^2$/s.

\begin{figure}[htbp]
    \centering
    \includegraphics{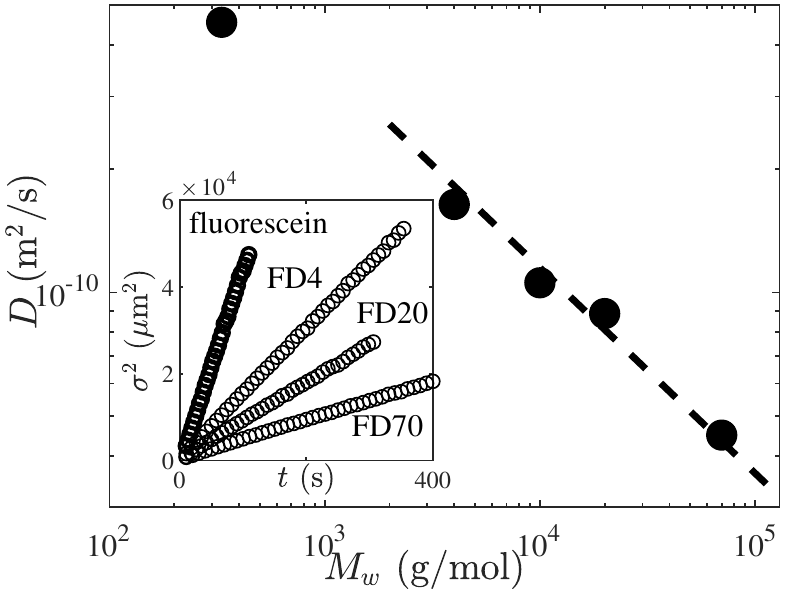}
    \caption{Measured diffusion coefficients $D$ vs  molecular weights $M_w$ of the solutes considered (fluorescein, dextran FD4, FD10, FD20, and FD70), see Table~\ref{table1}.
    The dotted line shows the scaling behavior $D \sim 1/\sqrt{M_w}$ for the dextrans.
    The inset shows the corresponding curves $\sigma^2$ vs\ $t$ from which $D$ are estimated by linear fits (the FD10 case is not shown for clarity).  
    \label{fig:CoeffDiff}}
\end{figure}

We performed these same measurements for dilute aqueous solutions of fluorescein and FD of higher molecular weights up to $M_w = 70$~kDa.
For all these experiments, Eq.~(\ref{eq:eqDiff}) correctly fits the 1D fluorescence intensity profiles and the linear behaviors $\sigma^2$ vs\ $t$ lead to precise estimates of the diffusion coefficients $D$, see 
the inset of Fig.~\ref{fig:CoeffDiff}. 
The corresponding $D$ values given in Table~\ref{table1} are also plotted 
in Fig.~\ref{fig:CoeffDiff} against $M_w$.
These values are in agreement with other  reported data measured  using the temporal widening of a fluorescein plug in a microchannel~\cite{Culbertson2002} and by modulated fringe pattern photobleaching for the FD solutions~\cite{Arrio-Dupont1996}.  The scaling law $D \sim M_w^{-1/2}$ shown in Fig.~\ref{fig:CoeffDiff} and also reported in Ref.~\cite{Arrio-Dupont1996}  is coherent with the random coil conformation of dextran molecules in water. \jb{Indeed, for
a loose polymer conformation, one expects that the
 radius of gyration  follows $R_g \sim \sqrt{M_w}$, and according to the Stokes-Einstein relation, $D \sim 1/R_h \sim 1/\sqrt{M_w}$ where $R_h \sim R_g$ is the hydrodynamic radius of the dextran coil in water~\cite{Arrio-Dupont1996}.} 
\begin{table}
\small
\caption{\label{table1} Molecular weight $M_w$ of the different solutes studied  and measured diffusion coefficients $D$ in water.}
\fontsize{7pt}{6pt}
\selectfont
\begin{ruledtabular}
\begin{tabular}{cccccccc}
Solutes & Fluorescein & FD4 & FD10 & FD20 & FD70\\
\hline
\hline
 $M_w$~(g/mol) & 332 & 4000 & 10000 & 20000 & 70000 \\
$D \times 10^{10}$ (m$^{2}$/s) & 4.5 & $1.6$ & $1.0$ & $0.9$ &  $0.46$ 
\end{tabular}
\end{ruledtabular}
\end{table}

\subsection{Evidence of diffusio-phoresis and diffusio-osmosis \label{sec:evidenceof}}

\begin{figure}[htbp]
    \centering
    \includegraphics{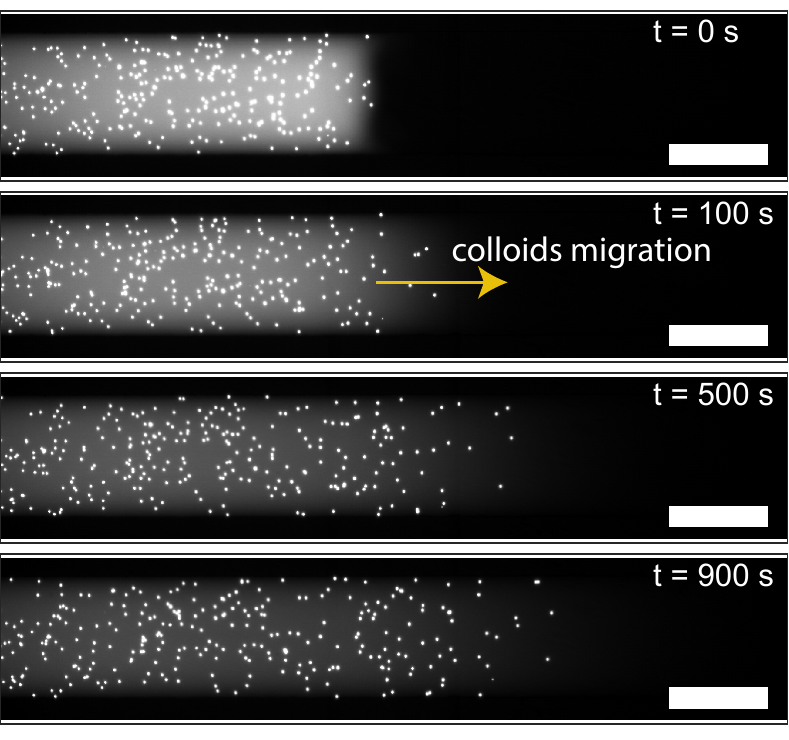}
    \caption{Fluorescence snapshots showing the temporal relaxation of a concentration gradient of FD70 with fluorescent colloids dispersed in the polymer solution. The colloids migrate
    to the low polymer concentration \jb{(see the arrow), because of diffusio-phoresis induced by the polymer concentration gradient, see Sec.~\ref{sec:DPonly}}.  
    Scale bar: $100~\mu$m (multimedia view).  \label{fig:DriftDP}}
\end{figure}

The results presented above, in particular the correct description of the temporal widening of the concentration gradient by Eq.~(\ref{eq:eqDiff}) with $\sigma = \sqrt{D t}$, suggest that the  solute transport is governed by diffusion only.
To confirm this hypothesis, we added fluorescent colloids (diameter $2 a = 2~\mu$m) to the FD70 solution and tracked their positions, $x_c$ vs $t$, along with the relaxation of the polymer concentration gradient.   
Strikingly, these experiments reveal that the colloids, in addition to their Brownian motion, move  towards the low polymer concentration, see Fig.~\ref{fig:DriftDP} for an experiment at $\varphi_0 = 0.005$ (multimedia view).

\begin{figure}[htbp]
    \centering
    \includegraphics{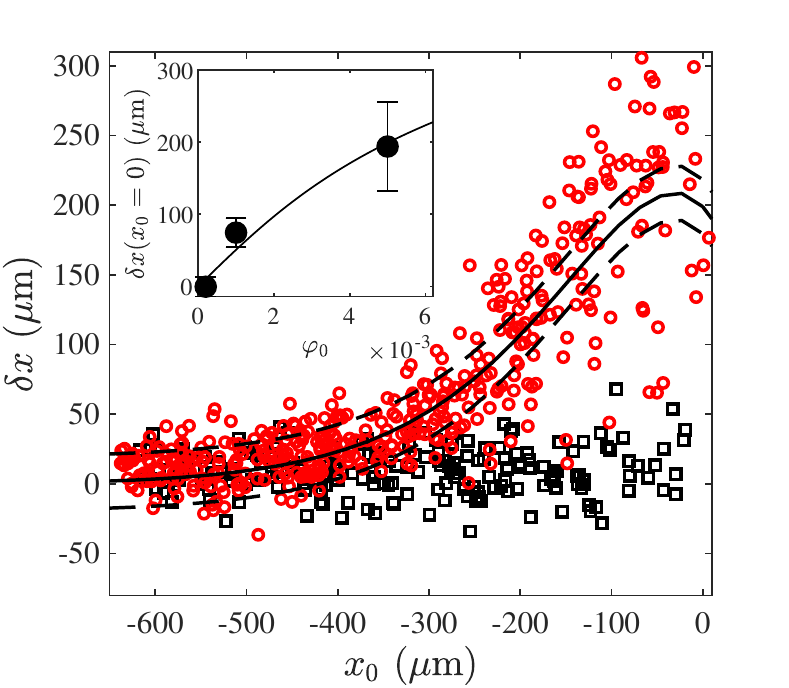}
    \caption{Colloid displacements $\delta x$ at $t = 840~$s  vs their initial positions $x_0$. Red circles: mutual diffusion  of water and  FD70 at concentration $\varphi_0° = 0.005$ with dispersed colloids;  
    black squares: colloid displacements in the absence of FD
    (diffusion in pure water).  The continuous line is the prediction by the solution of Eq.~(\ref{eq:dxcdt}). The two dashed lines show the Brownian dispersion $\pm \sqrt{2 D_\text{SD} t}$. 
    Inset: displacement $\delta x$ 
    at $t = 840$~s 
    for the colloids initially located at $x_0 \simeq 0$  
    vs\ initial volume fraction 
    $\varphi_0$ of the FD70  solution. 
    The errorbars are the standard deviations of the data. The solid line is the prediction given 
    by Eq.~(\ref{eq:xc}).}   
    \label{fig:DiffusioPhorese}
\end{figure}

To better visualize these displacements, Fig.~\ref{fig:DiffusioPhorese} shows the relative displacement at $t= 840$~s of each colloid from its initial position $\delta x = x_c(t)-x_0$ as a function
of $x_0 = x_c(t=0)$. 
At this time, the width of the polymer diffusion zone is $\sigma = \sqrt{D t} \simeq 200~\mu$m. We have retained in this analysis only colloids initially away from the channel edges ($\| y \| < w/2-20~\mu$m), in order to avoid possible confinement effects related to the rounded shape of the channel.
These data help to reveal that the colloid movement is correlated to the polymer gradient: particles initially upstream of the initial concentration gradient ($x_0 < -\sigma$) show very little displacements,  while particles initially close to $x =0$ show displacements that can reach $\delta x \simeq 250~\mu$m. 
Figure~\ref{fig:DiffusioPhorese} also shows the same data but for colloids dispersed in pure water only, showing in this case $\delta x \simeq 0$ with deviations that can be explained  by  Brownian motion, $\sqrt{D_{\text{SD}} t} \simeq  15~\mu$m, with
$D_{\text{SD}} \simeq 2.2\times 10^{-13}$~m$^2$/s the colloid self-diffusion coefficient estimated using the Stokes-Einstein relation. 


These observations  point  to the phenomena of diffusio-phoresis (DP) and diffusio-osmosis (DO). These  interfacial fluid transport phenomena are  due to specific interactions between the solute (FD polymers) and solid surfaces, that occur in a diffuse layer of small thickness either at the colloid surface or the channel walls~\cite{Derjaguin1972,Anderson1989,Marbach2019}. The solute concentration gradient induces a pressure gradient in the diffuse layer (parallel  to the interface) which is mechanically balanced by a viscous stress leading either to the net movement of freely suspended colloids (DP) or to a bulk flow  in the channel (DO).   
Such phenomena have been reported many times, in particular for electrolytes as solutes~\cite{Prieve1987,Abecassis2008,Nery-Azevedo2017,Shin2016,Ault2018,Alessio2021,Gu2018}, with a wide range of possible applications~\cite{Marbach2019,Velegol2016}, including microfluidics~\cite{Abecassis2009,Palacci2010,Shi2016,Shin2017}.  In the case of neutral solutes, small molecules but also polymers~\cite{Paustian2015,Lechlitner2018,Williams2020,Staffeld1989,Lee2014,Lee2017}, studies are scarcer and comparisons with theoretical models still raise issues~\cite{Lee2017,Williams2020}. \edi{This last case is also key to the understanding of other phenomena such as the stratification observed during the drying of colloidal films~\cite{Sear2017,Zhou2017,Schulz2020,Rees-Zimmerman2021,Schulz2022}}.

Our aim here is not to study these phenomena as such, but to show that the measured colloid displacements are consistent with DP induced by the dextran gradient and that DO probably also occurs simultaneously, explaining the deviations of the data shown in Fig.~\ref{fig:DiffusioPhorese}. 
We also show below that diffusio-osmotic flows do not a priori impact the $D$ measurements presented in Sec.~\ref{sec:molsol}. 

\subsubsection{Diffusio-phoresis induced by the dextran concentration gradient \label{sec:DPonly}}
For simplicity, we assume the Asakura-Oosawa model for the FD solution~\cite{Staffeld1989,Sear2017}: (i) no polymer-polymer interactions at the volume fractions $\varphi$ considered and (ii) polymer chains excluded from the colloid surface on a length scale $R_i \ll a$, therefore behaving as "hard-spheres". In this model, the DP drift velocity of a colloidal particle in the FD gradient is given by:
\begin{eqnarray}
v_\text{DP} = -\frac{ R_i^2 k_B T}{2 \eta_w}  \nabla n\,,
 \label{eq:DP}
 \end{eqnarray}
 with $k_B$ the Boltzmann constant and $n$ the number density of  polymer chains ($\#/$m$^3$).
For the FD70 solution, $n$ is estimated by
$n  = \varphi / \bar{\nu}_d$ 
with $\bar{\nu}_d  = M_w \nu_d/ \mathcal{N}_a$ the molecular volume of FD, $\mathcal{N}_a$ the Avogadro number, $M_w = 70~$kg/mol, and $\nu_d = 6.26 \times 10^{-4}$~m$^3$/kg the specific volume in Eq.~(\ref{eq:nu}).  This is obviously a rough estimate that ignores, for example, the molecular weight distribution of the polymer sample.

Following Eq.~(\ref{eq:eqDiff}) and Eq.~(\ref{eq:DP}), the trajectory of a colloidal particle $x_c$ vs\ $t$ due to DP when neglecting its Brownian motion is given by the solution of the ordinary differential equation:
\begin{eqnarray}
\frac{\text{d} x_c}{\text{d} t } = v_\text{DP}(x_c(t),t) =  \frac{R_i^2   k_B T \varphi_0}{2 \eta_w  \bar{\nu}_d} \frac{e^{-\frac{x_c^2}{4 D t}}}{\sqrt{4 \pi  D t}}\,,  \label{eq:dxcdt}
\end{eqnarray}
with $\varphi_0$ the volume fraction of the  initially injected FD solution.
With the dimensionless variables $\tilde{x}_c = x_c/h$ and $\tilde{t} = D t/h^2$, Eq.~(\ref{eq:dxcdt}) becomes 
\begin{eqnarray}
\frac{\text{d} \tilde{x}_c}{\text{d} \tilde{t}  }  =
 \varphi_0 \Gamma   \frac{e^{-\frac{\tilde{x}_c^2}{4 t}}}{\sqrt{\tilde{t} }}\,, \label{eq:traj}
\label{eq:dxcdtunitless}
\end{eqnarray}
with 
\begin{eqnarray}
\Gamma  = \frac{3 \sqrt{\pi}}{2}  \frac{R_\text{SE} R_i^2}{\bar \nu_d}\,,
\label{eq:Gamma}
\end{eqnarray}
and $R_\text{SE} \simeq 4.9$~nm the Stokes-Einstein radius  of the FD70 polymer particles defined  by $D = k_B T/(6\pi \eta_w R_\text{SE})$.
In the model of spherical "hard-sphere" colloids  considered in Ref.~\cite{Sear2017}, $R_i=R_\text{SE}$ and  $\bar \nu_d = 4 \pi R_\text{SE}^3/3$ so that $\Gamma = 9/(8\sqrt{\pi}) \simeq 0.63$. In this case, Eq.~(\ref{eq:dxcdtunitless}) and
the assumption  of dilute solution $\varphi_0 \ll 1$
necessarily impose a small DP velocity.

In the FD70 case shown in Fig.~\ref{fig:DiffusioPhorese}, however, 
the best fit of the experimental data  is obtained by the numerical solutions of Eq.~(\ref{eq:dxcdt}) for $R_i = 27$~nm, 
and thus for a relatively large interaction radius compared to $R_\text{SE}$ ($\Gamma \gg 1$).
Such a value is in agreement with previously reported data by Staffeld and Quinn who found $R_i \simeq 38$~nm also for dextran in water ($M_w = 40~$kg/mol, $R_\text{SE}= 4.8$~nm)~\cite{Staffeld1989}.
The fact that $R_i$ is significantly larger than $R_\text{SE}$ was interpreted by Staffeld and Quinn as being due to the flexible conformation of the dextran coil in water~\cite{Staffeld1989}. As discussed later on, this large difference remains surprising and raises   questions, because other studies have reported $R_i \simeq R_\text{SE}$ for polyethylene glycol polymers (PEG) using a different experimental configuration: DO in nanochannels~\cite{Lee2014,Lee2017}.

Equation~(\ref{eq:dxcdtunitless}) also shows that the magnitude of DP a priori depends on the initial polymer concentration $\varphi_0$.  The inset of Fig.~\ref{fig:DiffusioPhorese} confirms this result by showing the relative displacement $\delta x$ at $t = 840~$s of colloids that were close to the initial concentration gradient at $t=0$ ($x_0 \in [-h;0]$) for three different initial volume fractions of FD70 solutions ($\varphi_0 = 2\times 10^{-4}$, $1 \times 10^{-3}$, and $5 \times 10^{-3}$). Moreover, Eq.~(\ref{eq:dxcdtunitless}) admits a  self-similar solution for the initial condition $x_0=0$, which leads to:
\begin{eqnarray}
x_c = \sqrt{2 W(2\varphi_0^2\Gamma^2)\, D t}\,, \label{eq:xc}
\end{eqnarray}
$W$ being the Lambert function. 
As shown in the inset of Fig.~\ref{fig:DiffusioPhorese}, Eq.~(\ref{eq:xc}) correctly fits the results with the same parameters as above ($R_i=27$~nm).  Equation~(\ref{eq:xc}) also shows that the colloids always  remain "slaves" of the polymer diffusion in such experiments, even for possibly high $\varphi_0$ or $\Gamma$ values (high $R_i$), because of the slow logarithmic increase of the Lambert function in Eq.~(\ref{eq:xc}). This specificity and the effective diffusive behavior given by Eq.~(\ref{eq:xc})
are somehow similar to results derived by Ab\'ecassis \textit{et al.}\ for colloid DP induced by salt concentration gradients in coflow microfluidic experiments~\cite{Abecassis2008,Abecassis2009}.  
Because the  colloid dynamics  is  slaved to  solute diffusion, these  experiments are not the most appropriate for quantitatively studying DP transport, in particular Eq.~(\ref{eq:DP}),  compared to experiments that track colloid trajectories  in an independently imposed steady solute gradient as, for instance, in Refs.~\cite{Gu2018,Paustian2015,Shi2016,Williams2020}.

\subsubsection{Possible role of diffusio-osmosis}
Even if the aforementioned model roughly describes  our observations in quantitative agreement with previously reported data~\cite{Staffeld1989}, it cannot explain the large deviations in displacements shown in Fig.~\ref{fig:DiffusioPhorese}, which cannot be accounted for by the Brownian motion alone.
By changing the focal plane of the microscope during the relaxation of the polymer gradient, it appears that the colloid drift velocity also slightly depends on the altitude $z$ in the channel. 

\begin{figure}[htbp]
    \centering
    \includegraphics{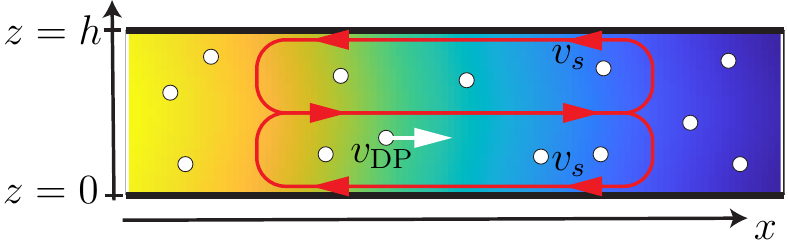}
    \caption{Superposition of DO (red) and colloid DP (white) in a channel of height $h$. The polymer concentration gradient is shown by the color gradient.     \label{fig:DODPFlow}}
\end{figure}

These observation possibly suggest the simultaneous development of DO within the channel due to the specific interaction of the polymer with the channel walls, as observed by other groups using  micro- and nanofluidic experiments for other neutral solutes, PEG~\cite{Lee2014,Lee2017} and glucose~\cite{Williams2020}. Indeed, if the  polymer chains are excluded from the channel walls on a given length scale, we also expect an apparent slip velocity $v_s$, given similarly to DP by Eq.~(\ref{eq:DP}), but with  opposite sign, the fluid flow being directed toward the high polymer concentration. 
Because of the volume continuity, these slip velocities induce a recirculating flow in the channel directly correlated with the polymer concentration gradient, even if the channel is closed by valve V$_1$, see Fig.~\ref{fig:DODPFlow} for a schematic cross-sectional view in the case of a 2D slit.
Colloids dispersed in the polymer solution are then subject to two transport mechanisms  as already shown in a similar microfluidic configuration (diffusion in dead-end pores) but for electrolyte solutions~\cite{Shin2016,Ault2018,Alessio2021}: migration by DP  and advection by DO, and the resulting colloid drift velocity can then depend on $z$. 

Our data also suggest that DO and DP have the same order of magnitude, i.e., $v_s \simeq - v_{\text{DP}}$, see schematically Fig.~\ref{fig:DODPFlow}. Indeed, if DO dominated DP, then we would have observed colloids transported to both the high and low polymer concentration (depending on $z$, see Fig.~\ref{fig:DODPFlow}), whereas if  DP dominated DO, we would not have  observed the large dispersion of the data shown in Fig.~\ref{fig:DiffusioPhorese}. 
Only time-resolved confocal fluorescence microscopy (due to the transient relaxation of the concentration gradient) would allow us to probe quantitatively the 3D flow and thus disentangle the  contributions due to  DO and DP to colloid movements.
These  difficult measurements, recently done by Williams {\it et al.}~\cite{Williams2020}
to disentangle DO from buoyancy-driven convection in their case, 
are, however, beyond the scope of this  work. They remain, nevertheless, necessary for a quantitative understanding of these interfacial-driven transport phenomena, and  might explain in particular, why $R_i$ estimated in Sec.~\ref{sec:DPonly} but also in Ref.~\cite{Staffeld1989} by considering only DP is significantly larger than $R_\text{SE}$ for dextran, whereas other experiments  have reported $R_i \simeq R_\text{SE}$ for DO induced by PEG gradients in nanochannels~\cite{Lee2014,Lee2017}.     

Furthermore,  it is legitimate to ask whether the presence of DO in the channel, induced by the  polymer concentration gradient, may impact the transport of the polymer itself (by advection) in the diffusion experiments presented in Sec.~\ref{sec:molsol}. To address this question, 
we first consider for simplicity the case of a 2D slit of height $h$ as in  Fig.~\ref{fig:DODPFlow}. The rounded shape of the fluidic channel would indeed require numerical tools to compute the 3D flow.  We then assume that the polymer concentration gradient  at the walls ($z=0$ and $z=h$) induces a DO slip velocity $v_s = - v_\text{DP}$, with $v_\text{DP}$ given by Eq.~(\ref{eq:DP}). The P\'eclet number $\text{Pe} = v_s h /D$  allows to compare the polymer diffusion and its advection by DO. The $D$ measurements shown previously are a priori only valid when the polymer advection is negligible, i.e., when $\text{Pe} \ll 1$. To go beyond this scaling analysis and also take into account the dependence of $v_s$ with the polymer concentration gradient, we performed an  analysis similar to the Taylor-Aris  dispersion, see the review in Ref.~\cite{Young1991} on solute dispersion in shear flows. 
In this framework, we showed in Appendix~\ref{App1} using the lubrication approximation that the height-averaged polymer concentration  $\langle \varphi \rangle$ follows: 
\begin{eqnarray}
\frac{\partial \langle \varphi \rangle}{\partial t} = \frac{\partial}{\partial x} \left( D_\text{eff}  \frac{\partial \langle \varphi \rangle}{\partial x} \right)\,,
\label{eq:dispersion}
\end{eqnarray}
with 
\begin{eqnarray}
D_\text{eff}= D\left[1 +  \frac{2 \pi\Gamma^2 h^2}{105}
\left(\frac{\partial \langle \varphi \rangle}{\partial x}\right)^2\right]\,.
\label{eq:DeffDisp}
\end{eqnarray}
The dispersion term in Eq.~(\ref{eq:DeffDisp}) accounts for advection of the polymer by DO, the non-linearity coming from the dependence of the magnitude of DO with the polymer gradient unlike the classical Taylor-Aris problem. The same non-linearity occurs for the dispersion of a buoyant solute in a horizontal channel, as solutal free convection likewise depends on the magnitude of the concentration gradient, see, e.g., Ref.~\cite{Salmon2021}.
In our experiments, the highest expected dispersion ($\varphi_0 = 0.005$ and $\sigma \sim h$) is about:
\begin{eqnarray}
D_\text{eff} \simeq D\left(1 +  \frac{2 \pi\Gamma^2 \varphi_0^2}{105}\right) \simeq 1.03 D\,,
\label{eq:DeffDisp2}
\end{eqnarray}
and diffusion thus always dominates the relaxation of the polymer gradient despite DO in the channel.
The aforementioned relationship suggests, however, that dispersion of the polymer by DO generated itself by the polymer gradient may occur for a higher concentration, nevertheless challenging the assumption of dilute  solution.

\subsection{Colloidal dispersions  \label{sec:colloiddisp}}

We have demonstrated in Sec.~\ref{sec:molsol} that the microfluidic experiments shown in Fig.~\ref{fig:Protocol} can be used to measure molecular diffusion coefficients in the range $D  \in [0.5-5]\times 10^{-10}$~m$^2$/s. To show that the same methodology can also be used  to measure much smaller diffusion coefficients, we performed similar experiments, but for dilute dispersions of colloids of radius $a = 250$~nm. 

\begin{figure}[htbp]
    \centering
    \includegraphics{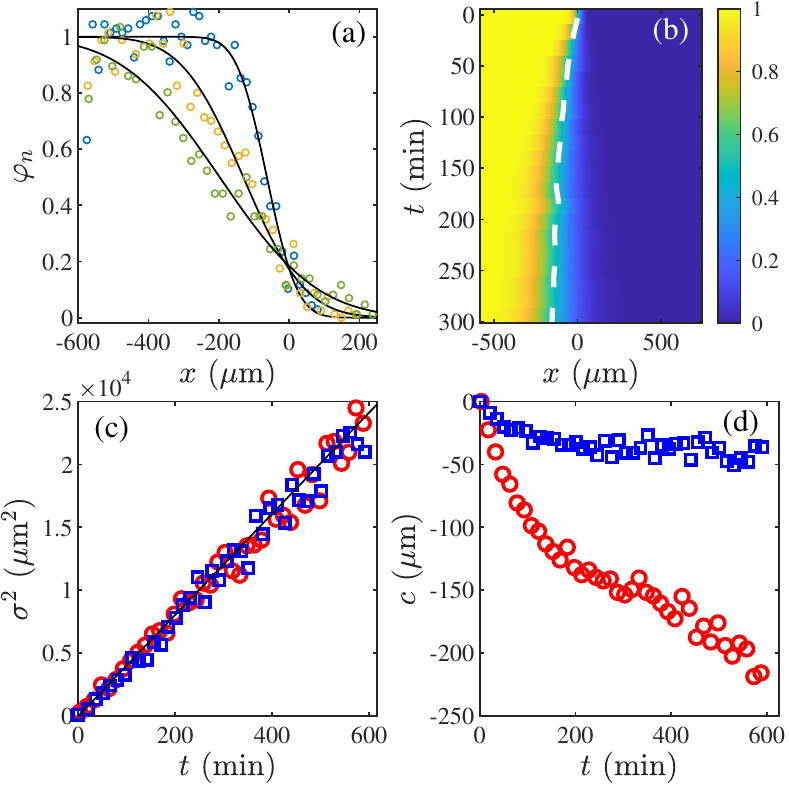}
    \caption{Colloid diffusion. (a) Experimental normalized concentration profiles $\varphi_n(x,t)$ at $t=63$, $213$, and $588$~min. The dark lines are fits by Eq.~(\ref{eq:eqDiffColl}).
    (b) Space-time plot of the fitted profiles for $L=4~$mm. The dotted line, $c$ vs $t$, shows the shift of the center of the diffusion zone. (c) $\sigma^2$ vs $t$ and (d) $c$ vs $t$  for two different
    experiments: $L=1~$mm (blue square) and $L=4~$mm (red circles). The linear fit in (c) gives
    $D \simeq 6.7 \times 10^{-13}$~m$^2$/s.
    \label{fig:CollectiveColloid}}
\end{figure}

To measure the 1D colloidal concentration field $\varphi(x,t)$, we studied dispersions of fluorescent colloids diluted enough with pure water  (typically $\simeq 10^{15}$~colloids/m$^3$) to be able to directly count the colloids and thus their density along the channel with  a good accuracy, see Fig.~\ref{fig:CollectiveColloid}(a).
As for the FD and fluorescein cases, these concentration profiles are normalized by 1, but Eq.~(\ref{eq:eqDiff}) does not capture the temporal relaxation of the colloid concentration gradient, see Fig.~\ref{fig:CollectiveColloid}(a). However, we found that the  relation:
\begin{equation}
 \varphi_n(x,t) = \frac{1}{2} \left[1-\text{erf}\left(\frac{x-c}{2\sigma}\right)\right]\,,
 \label{eq:eqDiffColl}
 \end{equation}
 correctly fits the data, $c$ and $\sigma$ being  free parameters. 
 The space-time diagram shown in Fig.~\ref{fig:CollectiveColloid}(b) corresponding to these fits clearly shows an overall convective motion, directed toward the high colloid concentrations, superimposed on the broadening of the mixing zone by diffusion. Figure~\ref{fig:CollectiveColloid}(c) shows that the width of the mixing zone $\sigma$ follows $\sigma^2 = D t$ leading to an estimate of the colloid diffusion coefficient $D \simeq 6.7 \times 10^{-13}$~m$^2$/s.
 For this experiment, the distance between the two valves (see Fig.~\ref{fig:Setup}) is $L=4$~mm and the data $c$ vs $t$ plotted in Fig.~\ref{fig:CollectiveColloid}(d) show a global displacement of the diffusion zone of  $\simeq 200~\mu$m in $\simeq 600$~min  that gradually slows down.
 
 We believe this very weak convective motion is due to colloid  DP  due to the relaxation of a molecular solute gradient present in minute amounts, simultaneously with colloid diffusion. Indeed, the commercial colloidal dispersion  contains traces of surfactants or ionic solutes such as sodium azide as a preservative, that can induce DP, toward the high solute concentration  in our experiments, as shown in many experiments reporting colloid DP in solutions of electrolytes~\cite{Abecassis2008,Shin2016}. 
 To test this assumption, we performed the same experiments but with a new chip for which the distance between the two valves is now $L=1$~mm. Again, the concentration profiles are well-fitted by Eq.~(\ref{eq:eqDiffColl}) and the width of the diffusion zone follows $\sigma^2 = D t$ with the same $D$ value, see Fig.~\ref{fig:CollectiveColloid}(c).  In this case, however, the shift $c$ vs $t$  reaches for $t \simeq 100~$min a small plateau value $c \simeq -40~\mu$m, see Fig.~\ref{fig:CollectiveColloid}(d). We believe that this quantitatively different behavior is due to the relaxation by diffusion of the concentration gradient of the molecular impurity responsible for the colloid DP. Indeed, the initial volume of the dispersion is defined in our protocol by the distance $L$ between the two valves, see Fig.~\ref{fig:Protocol}. The concentration gradient of the impurity initially contained in the dispersion  relaxes 
on a timescale $\sim L^2/D_i$ with $D_i$ its diffusion coefficient. For  $D_i \simeq 5 \times 10^{-10}$~m$^2$/s,   $L^2/D_i \simeq 30$~min for  $L= 1$~mm and is 16 times smaller than the $L=4~$mm case ($L^2/D_i \simeq 530$~min).
These simple arguments allow us to account, at least qualitatively, for the behaviors shown in Fig.~\ref{fig:CollectiveColloid}(d), and also show how the distance $L$ between the valves  can be adapted to avoid colloid DP (with the assumption $D \ll D_i$). 
Another possibility to get rid of  DP  would be to intentionally add salts at the same concentration to both the initial water and dispersion reservoirs to saturate any ionic impurity background initially present in the dispersion. In the case of concentrated colloidal dispersions, this strategy could on the other hand modify the colloidal interactions and thus the measured diffusion coefficient~\cite{Russel}.

 The aforementioned experiments lead to a precise estimate of the diffusion coefficient of the colloids 
$D \simeq 6.7 \pm 0.3 \times 10^{-13}$~m$^2$/s, the errorbar corresponding to the  deviation over several experiments. This value, estimated from the relaxation of a colloid  concentration gradient over long time scales ($\simeq 600$~min) is known as the mutual or long-time collective diffusion coefficient~\cite{Russel}.  For comparison, we also measured the self-diffusion coefficient $D_\text{SD}$ of the same colloids, tracking their mean square displacements on short time scales (a few seconds), see Appendix~\ref{App2}.
 These measurements lead to $D_\text{SD} \simeq 7.9 \pm 0.5 \times 10^{-13}$~m$^2$/s corresponding to a colloid radius $a \simeq 270~\pm 15$~nm using the Stokes-Einstein relation at $T=22^\circ$C (manufacturer data,  $a=250$~nm). For this dilute dispersion, we expect $D \simeq D_\text{SD}$ because both colloidal and hydrodynamic interactions are a priori negligible~\cite{Russel}. The slight measured difference ($\simeq 15 \%$) may be due to  our too rough estimation of $D_\text{SD}$ from the colloid mean square displacements (see Appendix~\ref{App2}), or to experimental artifacts in our measurements of the colloid concentration profiles due to the unavoidable adhesion of some colloids to the channel walls on long time scales.   This slight difference could also be due to more subtle effects such as the confinement of the colloids in the channel, especially near the edges due to its rounded shape which may modify their  self-diffusivity~\cite{Brenner1961}, \edi{see Ref.~\cite{Avni2021} for the case of charged colloids.}

\section{Conclusions}
In this work, we have developed a microfluidic chip to track the relaxation of a steep solute concentration gradient in a microchannel closed to suppress  convection.
In the case of dilute solutions and colloidal dispersions, this methodology allowed us to measure solute diffusion coefficients over a very wide range, $D \in [10^{-13}- 10^{-9}]$~m$^2$/s, without convection disturbance.  Due to the high control 
of the mass transport conditions at these small scales, these experiments also revealed  interfacial-driven transport phenomena, colloid DP and DO, due to the presence of the solute concentration gradient. We have nevertheless shown that these transport phenomena do not interfere with the $D$ measurements made under our conditions.
These experiments could in turn provide new data on these phenomena, especially in the case of neutral solute, if one is able to disentangle transport due to DP and DO using for instance time-resolved  confocal microscopy or chemical modifications of the channel walls or colloid surfaces to suppress either DP or DO.

The methodology described in the present work could also be useful for probing mutual (collective) diffusion in more concentrated systems, for which solute-solute and hydrodynamic interactions are expected to play a role~\cite{Russel}.  
In the case of charged colloidal dispersions containing both salts and colloids, being able to adjust the distance $L$ between the valves to decouple the diffusion of salts from that of colloids (see Sec.~\ref{sec:colloiddisp}) is a major asset not only for measuring the collective diffusion coefficient of dispersions in equilibrium with a reservoir of known  ionic content~\cite{Keita2021}, but also for tackling the complexity of mass diffusion in such systems~\cite{Annunziata2020}.
Finally,  the permeation property of the PDMS matrix of the chip could also be exploited to concentrate solutes in the channel by pervaporation~\cite{Bacchin2022}, prior the diffusion experiment triggered by the opening of valve V$_2$, see Fig.~\ref{fig:Protocol}. Such experiments would then allow to probe the phenomenon of mutual diffusion using the technique of microfluidic free interface diffusion, even in the case of highly concentrated solutions or dispersions. 

\appendix
\section{Solute dispersion by diffusio-osmosis \label{App1}}

We  consider  the experiments described in Sec.~\ref{sec:molsol} but for a 2D slit of height $h$. At the microfluidic scales ($h\simeq 23~\mu$m in the experiments), inertia is totally negligible and the  dispersion of the solute by free convection can also be neglected~\cite{Salmon2021}. The equations governing the polymer volume fraction and the flow are thus:
\begin{eqnarray}
&&\frac{\partial \varphi}{\partial t} + \mathbf{v}. \nabla \varphi = D \Delta \varphi\,,\\
&&\nabla.\mathbf{v} = 0\,,\\
&&\eta_w \Delta \mathbf{v} = \nabla p\,. 
\end{eqnarray}
where $\mathbf{v}$ is the flow due to DO induced by the polymer concentration gradient, and $p$ is the pressure field.  We assumed above 
that the polymer solution is dilute with a constant viscosity $\eta_w$. 
DO arises  because of the repulsion of the polymer chains from the walls of the slit, which results in the following  slip velocity:
\begin{eqnarray}
v_x(z=0~\text{and}~h) = \frac{k_B T R_i^2}{2 \eta_w \bar \nu_d} \left(\frac{\partial \varphi}{\partial x}\right)_{z=0,h}\,,
\end{eqnarray}
which is similar to Eq.~(\ref{eq:DP}) corresponding to DP of a colloidal particle in the polymer gradient.
With the  dimensionless
variables:
\begin{eqnarray}
&&\tilde{t} = D t/h^2,~ \tilde{x}=x/h,~ \mathbf{\tilde{v}}= h\mathbf{v}/D,\notag\\ 
&&\tilde{p}=h^2 p/ (\eta_w D),~ \tilde{\varphi}=\varphi/\varphi_0\,,
\end{eqnarray}
the aforementioned  model reads:
\begin{eqnarray}
&&\frac{\partial \tilde\varphi}{\partial \tilde t} + \mathbf{\tilde v}. \nabla \varphi = \Delta \tilde\varphi\,, \label{eq:aditransport}\\
&&\nabla.\mathbf{\tilde v} = 0 \label{eq:adicontinuity}\,,\\
&&\Delta \mathbf{\tilde v} = \nabla \tilde p\,,
\end{eqnarray}
and:
\begin{eqnarray}
\tilde v_x(\tilde z=0~\text{and}~\tilde z=1) = 2 \sqrt{\pi}\varphi_0\Gamma \left(\frac{\partial \tilde\varphi}{\partial \tilde x}\right)_{\tilde z=0,1}\,.
\end{eqnarray}
We now assume:
\begin{eqnarray}
\tilde\varphi(\tilde x,\tilde z,\tilde t) = \langle\tilde{\varphi}\rangle(\tilde x,\tilde t)+\tilde\varphi_1(\tilde x,\tilde z,\tilde t)\,,
\end{eqnarray}
with $\langle\tilde{\varphi}\rangle$ the height-averaged concentration defined by:
\begin{eqnarray} \langle\tilde{\varphi}\rangle= \int_0^1 \tilde \varphi(\tilde x, \tilde z,\tilde t) \text{d}\tilde z\,,
\end{eqnarray}
and $\tilde\varphi_1 \ll \langle\tilde{\varphi}\rangle$.
Averaging of the transport equation  Eq.~(\ref{eq:aditransport}) over the height of the slit  with the help of Eq.~(\ref{eq:adicontinuity}) leads to:
\begin{eqnarray}
\frac{\partial \langle\tilde{\varphi}\rangle}{\partial \tilde t} 
+ \frac{\partial \langle \tilde v_x \tilde\varphi_1 \rangle}{\partial  \tilde x}
= \frac{\partial^2 \langle\tilde{\varphi}\rangle}{\partial \tilde x^2}\,.
\label{eq:transaveraged}
\end{eqnarray} 
In the framework of the lubrication approximation, i.e., for a polymer concentration gradient extended over a length scale~$\gg 1$, one can show 
by substracting Eq.~(\ref{eq:transaveraged}) to the transport equation Eq.~(\ref{eq:aditransport}) that we get: 
 \begin{eqnarray}
 \tilde{v}_x\frac{\partial \langle\tilde{\varphi}\rangle}{\partial \tilde x} \simeq \frac{\partial^2 \tilde\varphi_1}{\partial \tilde z^2}\,,   \label{eq:intermediaire}
\end{eqnarray}
for $t \gg 1$,  see  Ref.~\cite{Young1991} on solute dispersion in shear flows for more details on such calculations. 
Similarly, it can be shown, still within the lubrication approximation, that the $x$-component of the velocity field is given by:
\begin{eqnarray}
 \tilde{v}_x \simeq 2\sqrt{\pi}\varphi_0\Gamma \frac{\partial \langle\tilde{\varphi}\rangle}{\partial \tilde x} (6 \tilde z^2 - 6 \tilde z +1)\,.\label{eq:vfluidadi}
 \end{eqnarray}
 Equation~(\ref{eq:intermediaire}) along with Eq.~(\ref{eq:vfluidadi}) and the impermeability conditions at the walls can be used to compute $\tilde\varphi_1$: 
 \begin{eqnarray}
\tilde\varphi_1 =  2 \sqrt{\pi}\varphi_0\Gamma\left(\frac{\partial \langle\tilde{\varphi}\rangle}{\partial \tilde x}\right)^2
 \left(\frac{\tilde z^4}{2}-\tilde z^3+\frac{ \tilde z^2}{2}-\frac{1}{60}\right).\label{eq:phiadi}
\end{eqnarray}
 Equations~(\ref{eq:phiadi}) and~(\ref{eq:vfluidadi})  can then used to calculate  the dispersion term 
 in Eq.~(\ref{eq:transaveraged})  leading to  Eq.~(\ref{eq:dispersion}) with real units.  Similar calculations are also derived and discussed in detail in Ref.~\cite{Salmon2021} dealing with solute dispersion due to free convection in a microfluidic slit.

\section{Colloid self-diffusion coefficient \label{App2}}

\begin{figure}[htbp]
    \centering
    \includegraphics{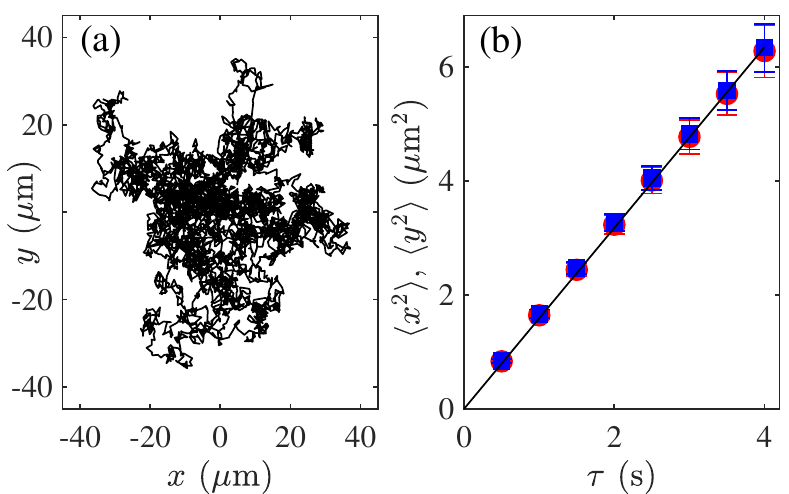}
    \caption{(a) Example of a colloid trajectory in the $x$-$y$ plane. (b) Mean square displacement
    of the colloids vs time $\tau$ (blue squares: $x$-direction, red bullets: $y$-direction). The errorbars are estimated from the deviation over $\simeq 50$ different trajectories. The solid line is the best fit by $\langle x^2 \rangle = \langle y^2 \rangle = 2 D_\text{SD} \tau$ 
    \label{fig:FgMSD.pdf}}
\end{figure}

To estimate the self-diffusion coefficient of the colloids, we first filled the entire microfluidic channel with the dispersion used in the experiments described in Sec.~\ref{sec:colloiddisp} at a higher dilution ($\simeq 5 \times 10^{13}$~colloids/m$^3$) and then closed valve~V$_1$. 
We then acquired images for $\simeq 1$~h on a field of view of $\simeq w \times w$, $w$ being the channel width.
Two-dimensional trajectories of the colloids were extracted from the image stack using standard particle tracking algorithms~\cite{PTcode}, see Fig.~\ref{fig:FgMSD.pdf}(a).
From about 50  trajectories, we then computed the mean square displacement $\langle x^2 \rangle$ and $\langle y^2 \rangle$ against time $\tau$.  Figure~\ref{fig:FgMSD.pdf}(b) shows that  
$\langle x^2 \rangle \simeq \langle y^2 \rangle = 2 D_\text{SD} \tau$ with $D_\text{SD} \simeq 7.9 \pm 0.5 \times 10^{-13}$~m$^2$/s, the standard deviation being calculated from the deviation over the different trajectories. 

These measurements of the self-diffusion coefficient would deserve a finer estimation and more discussion~\cite{PCM}, notably on the number of trajectories followed, the precision on the localization of the particles, the statistical estimation of $D_\text{SD}$ from the trajectories~\cite{Vestergaard2014}, the size distribution of the tracers, the role of 
confinement~\cite{Brenner1961}, see also Ref.~\cite{Avni2021} for the case of charged colloids, etc.

\begin{acknowledgments}
We thank C. Ybert for his insightful comments. We also  acknowledge Solvay, CNRS and the ANR program Grant No.~ANR-18-CE06-0021 for financial support. \end{acknowledgments}

%

\end{document}